\documentclass[useAMS,usenatbib]{mn2e}
\usepackage[dvips]{graphicx}
\usepackage[english]{babel}
\usepackage{amsmath}
\usepackage{amssymb}
\usepackage{textcomp}
\usepackage{gensymb}
\usepackage{natbib}

\title[The evolution of a misaligned gas disc]{The creation and persistence of a misaligned gas disc in a simulated early-type galaxy}
\author[F. van de Voort et al.]{Freeke~van~de~Voort$^{1,2}$\thanks{E-mail: freeke@berkeley.edu}, 
Timothy~A.~Davis$^3$, Du\v{s}an~Kere\v{s}$^4$, Eliot~Quataert$^1$, 
\newauthor
Claude-Andr\'e~Faucher-Gigu\`ere$^5$, and Philip~F.~Hopkins$^6$ \\ 
$^{1}$Department of Astronomy and Theoretical Astrophysics Center, University of California, Berkeley, CA 94720-3411, USA \\
$^{2}$Academia Sinica Institute of Astronomy and Astrophysics, P.O. Box 23-141, Taipei 10617, Taiwan \\
$^{3}$Centre for Astrophysics Research, University of Hertfordshire, Hatfield, Herts AL1 9AB, UK \\
$^{4}$Department of Physics, Center for Astrophysics and Space Sciences, University of California at San Diego, 9500 Gilman Drive, \\
\ \ La Jolla, CA 92093 \\
$^{5}$Department of Physics and Astronomy and CIERA, Northwestern University, 2145 Sheridan Road, Evanston, IL 60208, USA \\
$^{6}$TAPIR, Mailcode 350-17, California Institute of Technology, Pasadena, CA 91125, USA 
}

\begin{document}

\date{Accepted May 27, 2015. Received May 26, 2015; in original form April 14, 2015}

\pagerange{\pageref{firstpage}--\pageref{lastpage}} \pubyear{2015}

\maketitle

\label{firstpage}

\begin{abstract}

Massive early-type galaxies commonly have gas discs which are kinematically misaligned with the stellar component. These discs feel a torque from the stars and the angular momentum vectors are expected to align quickly. We present results on the evolution of a misaligned gas disc in a cosmological simulation of a massive early-type galaxy from the Feedback In Realistic Environments project. This galaxy experiences a merger which, together with a strong galactic wind, removes most of the original gas disc. The galaxy subsequently reforms a gas disc through accretion of cold gas, but it is initially 120\degree\ misaligned with the stellar rotation axis. This misalignment persists for about 2~Gyr before the gas-star misalignment angle drops below 20\degree.  The time it takes for the gaseous and stellar components to align is much longer than previously thought, because the gas disc is accreting a significant amount of mass for about 1.5~Gyr after the merger, during which the angular momentum change induced by accreted gas dominates over that induced by stellar torques. Once the gas accretion rate has decreased sufficiently, the gas disc decouples from the surrounding halo gas and realigns with the stellar component in about 6 dynamical times. During the late evolution of the misaligned gas disc, the centre aligns faster than the outskirts, resulting in a warped disc. We discuss the observational consequences of the long survival of our misaligned gas disc and how our results can be used to calibrate merger rate estimates from observed gas misalignments.

\end{abstract}


\section{Introduction} \label{sec:intro}

Elliptical and lenticular galaxies are thought to form at high redshift through repeated mergers of smaller galaxies (e.g. \citealt{Toomre1977}). Such `early-type galaxies' (ETGs) are generally among the most massive galaxies in our universe. Local ETGs are mostly gas free and dominated by old stars, while lower mass spiral galaxies have a more massive interstellar medium (ISM) and are still forming stars. This anti-hierarchical growth of luminous structure (termed `downsizing'; \citealt{Cowie1996}) is one of the key puzzles in an otherwise hierarchical cold dark matter (CDM) universe. 

The gas-poor nature of ETGs makes them excellent laboratories in which to study the effects that stellar mass-loss, cooling from the hot halo, galaxy mergers, and accretion have on galaxy evolution \citep[e.g.][]{Kaviraj2014}.
Studies of statistically complete samples of ETGs have shown that about a quarter of these objects have more than 10$^7$~M$_{\odot}$ of cold molecular gas \citep{Combes2007, Welch2010, Young2011}  and approximately 40 per cent have sizeable atomic gas reservoirs \citep{Morganti2006, Sage2006, Serra2012}. Dust is also present in about 50 per cent of ETGs \citep[e.g.][]{Smith2012}. This ISM material appears to be regenerating previously gas-poor galaxies \citep{Young2014}, but its origin is debated.

Observationally, in field environments the gas and dust in ETGs seems to come mainly from external sources, such as mergers and cold gas accretion \citep[e.g.][]{Knapp1989, Sarzi2006, McDermid2006, Oosterloo2010, Davis2011, Smith2012, Duc2015}.
Theoretically, however, it has been argued that the main mode of gas supply is cooling from hot halo gas \citep{Lagos2014, Lagos2015}, with minor mergers playing a secondary role. These semi-analytic models had to make assumptions about some of the relevant physical processes, which need to be verified by hydrodynamic simulations. 

Here we concentrate on understanding the kinematic decoupling which has been observed between the gaseous and stellar components of ETGs \citep[e.g.][]{Sarzi2006, Davis2011, Katkov2014}. 
About 50 per cent of local field ETGs have gas kinematically misaligned from their stellar kinematic axis \citep{Davis2011, Serra2014}. There are indications of the same behaviour at high redshift. At $z\approx1-2.5$, \citet{Wisnioski2015} found large misalignments between the gas kinematic and stellar photometric axis in several galaxies. These galaxies tended to have redder colours and lower specific star formation rates than the majority of galaxies in their sample, consistent with these objects being ETGs\footnote{Star-forming galaxies have much higher gas fractions than ETGs and are not likely to exhibit significantly misaligned gas discs, because they are unlikely to accrete enough material (and angular momentum) to tilt the entire gas disc.} 

A priori, it seems likely that that the origin of misaligned material has to be external, as gas produced by internal processes (e.g. stellar mass-loss) should share the angular momentum of the stars \citep{Davis2011}. 
Thus the observed incidence of misalignments gives a lower limit for the fraction of galaxies where the gas must have been supplied by external processes. 
However, in an axisymmetric potential the gas will over time relax back into one of the preferred axes, becoming exactly co- or counter-rotating. 
This is because the gas feels a torque in the (quadrupolar) gravitational potential of the stars. This torque will make the gas disc precess around the angular momentum direction of the stars. Because the torque is larger at smaller radii, gas will precess at a faster rate. Due to the fact that the gas is not on exactly circular orbits, there will be collisions of gas clouds precessing at different rates. These collisions work to realign the differentially precessing gas `rings'. Dissipative forces destroy the component of the angular momentum of the gas perpendicular to the stellar angular momentum direction, until they are aligned or 180\degree\ misaligned, the most stable configurations. 

Theoretical studies of the relaxation of misaligned gas discs in the potential of elliptical galaxies have considered this effect in both axisymmetric and triaxial halos and found that in both cases the alignment process typically takes a few dynamical times \citep[$t_\mathrm{dyn}$;][]{Tohline1982, Lake1983}. \cite{Lake1983} found the relaxation time, i.e.\ the time it takes a misaligned disc to settle into the plane, to be similar to the differential precession time, which is equal to
$t_\mathrm{relax} \approx t_\mathrm{dyn}/\epsilon$, where $\epsilon$ is the eccentricity of the potential. For a typical lenticular galaxy $\epsilon\approx0.2$ \citep{Mendez2008}, and $t_\mathrm{relax}\approx5\, t_\mathrm{dyn}$.
This time is \textit{very} short in the centre of local ETGs (about 100 Myr), and thus very few misaligned objects should be observed. It is this puzzling mismatch between the theoretical predictions of fast relaxation and observations of many misaligned discs that we aim to investigate here. 
Davis \& Bureau (in preparation) point out that if one can understand the time this relaxation process takes, one can obtain (at low redshift, where mergers are expected to dominate the supply of cold gas) an estimate of the rate of gas-rich mergers in massive galaxies, a quantity which is important to galaxy evolution in a standard $\Lambda$CDM framework.

In this paper we conduct a case study, investigating the evolution of a gas disc embedded in the centre of a large ETG within a cosmological, hydrodynamic simulation. This gas disc is accreted in a misaligned configuration and relaxes slowly back towards the stellar kinematic axis as the galaxy evolves. We find that the gas-star misalignment persists for more than 2~Gyr, much longer than previously expected, due to continuous gas accretion. In Section~\ref{sec:sim} we describe the simulation we used. In Section~\ref{sec:results} we present our results, with Section~\ref{sec:prop} focussing on the galaxy's accretion history and metallicity and Section~\ref{sec:align} focussing on the gas disc's misalignment and relaxation. The main results are shown in Figures~\ref{fig:angle} and~\ref{fig:relax}. We discuss our results and conclude in Section~\ref{sec:concl}.

\section{Method} \label{sec:sim}

This work is part of the `feedback in realistic environments' (FIRE) project\footnote{http://fire.northwestern.edu/} \citep{Hopkins2014FIRE}, which consists of several cosmological `zoom-in' simulations. Here we make use of one galaxy, \textbf{m13}, with stellar mass $M_\mathrm{star}=10^{11}$~M$_\odot$,  similar to that of giant ETGs at $z=0$, and with a halo mass\footnote{We define halo mass as the total mass inside a spherical region within which the mean density is 200 times the mean density of the Universe} $M_\mathrm{halo}=10^{13}$~M$_\odot$. Its initial conditions are taken from the AGORA project \citep{Kim2014}. The simulation is fully described in \citet{Hopkins2014FIRE} and references therein, but we will summarize its main properties here.
 
We use a version of TreeSPH (P-SPH) which adopts the Lagrangian `pressure-entropy' formulation of the smoothed particle hydrodynamics (SPH) equations \citep{Hopkins2013PSPH}. The gravity solver is a heavily modified version of \textsc{gadget}-2 \citep[last described in][]{Springel2005}. P-SPH also includes substantial improvements in the artificial viscosity, entropy diffusion, adaptive timestepping, smoothing kernel, and gravitational softening algorithm \citep{Hopkins2014FIRE}.

A $\Lambda$CDM cosmology is assumed with parameters consistent with the 9-yr Wilkinson Microwave Anisotropy Probe (WMAP) results \citep{Hinshaw2013}, $\Omega_\mathrm{m}= 1-\Omega_\Lambda = 0.272$, $\Omega_\mathrm{b}= 0.0455$, $h = 0.702$, $\sigma_8 = 0.807$ and $n = 0.961$.
The (initial) particle masses for dark matter and baryons are $2.3\times10^6$~M$_\odot$ and $4.5\times10^5$~M$_\odot$, respectively, for our fiducial resolution. The minimum physical baryonic force softening length for our fiducial resolution is 28~$h^{-1}$pc. We adopt a quintic spline kernel with an adaptive size, keeping the mass within the kernel approximately equal. The average number of neighbours in the smoothing kernel is 62. 
The contamination fraction, i.e.\ the fraction of the total dark matter mass that is in lower resolution (higher mass) particles, at $z=0$ is 0.07 per cent within 20~kpc, the region we study. Therefore, we do not expect contamination to affect our results. 

Star formation takes place in molecular, self-gravitating gas above a hydrogen number density of $n_\mathrm{H}^\star>10$~cm$^{-3}$, where the molecular fraction is calculated following \citet{Krumholz2011} and the self-gravitating criterion following \citet{Hopkins2013SelfGrav}. Stars are formed at the rate $\dot\rho_\star=\rho_\mathrm{molecular}/t_\mathrm{ff}$, where $t_\mathrm{ff}$ is the free-fall time. Star particles inherit their mass and metal abundances from their progenitor gas particle.
We assume an initial stellar mass function from \citet{Kroupa2002}. Radiative cooling and heating are computed in the presence of the cosmic microwave background (CMB) radiation and the ultraviolet/X-ray background from \citet{Faucher2009}. Self-shielding is accounted for with a local Sobolev/Jeans length approximation. We impose a temperature floor of 10~K or the CMB temperature. 

The primordial abundances are $X = 0.76$ and $Y = 0.24$, where $X$ and $Y$ are the mass fractions of hydrogen and helium, respectively. The simulation has a metallicity floor at $Z=10^{-4}$~Z$_\odot$, because yields are very uncertain at lower metallicities. The abundances of 11 elements (H, He, C, N, O, Ne, Mg, Si, S, Ca and Fe) produced by massive and intermediate-mass stars are computed following \citet{Iwamoto1999}, \citet{Woosley1995}, and \citet{Izzard2004}. Mass ejected by a star particle through stellar winds and Type Ia and Type II supernovae is transferred to the gas particles in its smoothing kernel.

The FIRE simulations include an implementation of stellar feedback by supernovae, radiation pressure, stellar winds, and photo-ionization and photo-electric heating (see \citealt{Hopkins2014FIRE} and references therein for details). For the purposes of the present paper, we emphasize that these simulations produce galaxies with stellar masses and metallicities reasonably consistent with observations over a wide range of dark matter halo masses \citep{Hopkins2014FIRE, Ma2015}. This is a consequence of galactic winds driven by stellar feedback \citep[see also][]{Muratov2015}. Feedback from active galactic nuclei is not included. 

All distances in this work are given in proper~kpc. Many results are shown as a function of lookback time, $t_\mathrm{lookback}$. In our cosmology, $t_\mathrm{lookback}=0$, 1, 2, 3, 4, and 5~Gyr correspond to $z=0$, 0.075, 0.16, 0.26, 0.37, and 0.49, respectively. A major galaxy merger occurs at $t_\mathrm{lookback}=3.4$~Gyr ($z=0.3$), which is indicated in the figures by the grey shaded region.

\section{Results} \label{sec:results}

\begin{figure*}
\center
\includegraphics[scale=.52]{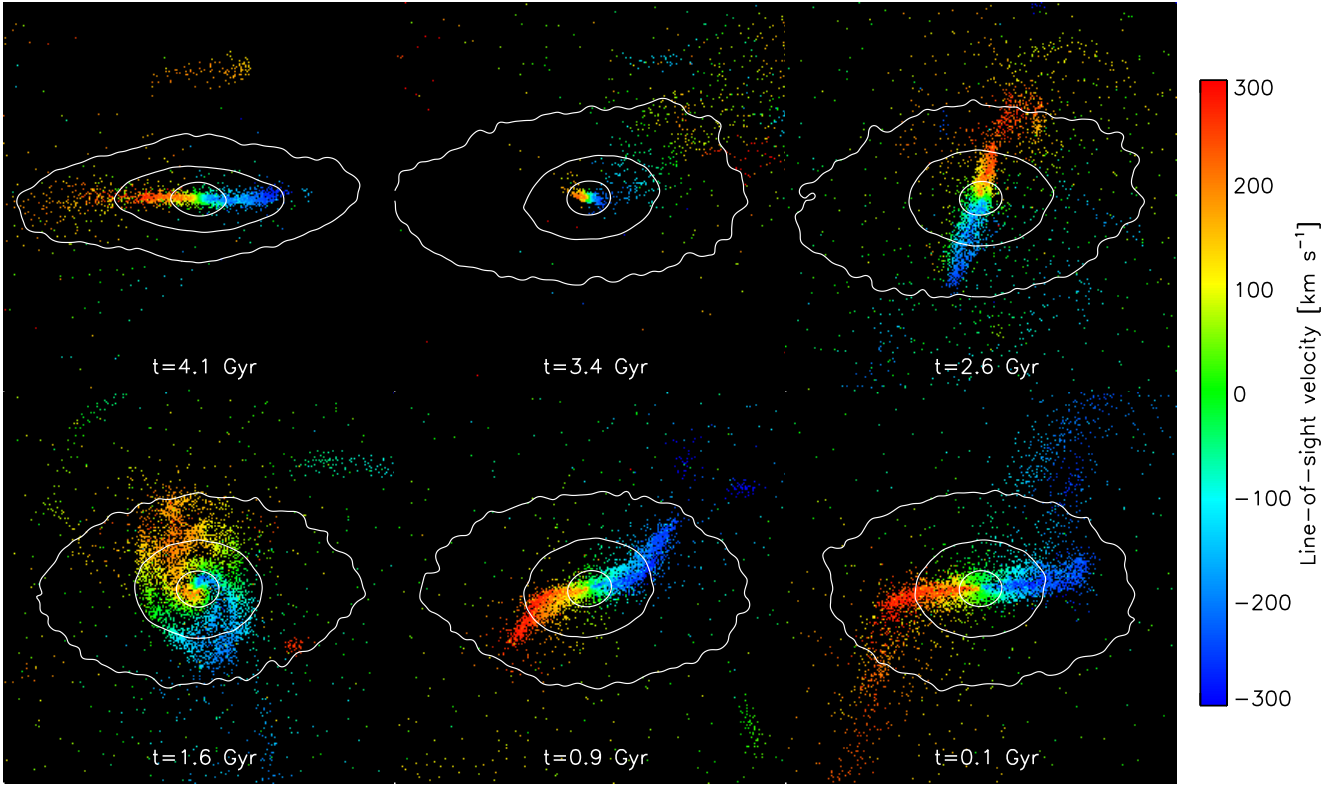}
\caption {\label{fig:img} The average line-of-sight velocity of all gas particles in a (20~kpc)$^3$ box centred on the galaxy at several lookback times, $t$, as indicated in each panel. The colour scale runs from $-300$ (blue) to 300~km~s$^{-1}$ (red) and is saturated at either end. The image is rotated such that the stellar angular momentum vector points in the $y$-direction (i.e.\ upwards). The shape of the stellar component is shown as white contours which connect areas of equal mass surface density, with logarithmically spaced contour levels. The first panel shows the gas disc before the merger, the second panel right after the merger when most of the gas disc has been destroyed. The galaxy reforms a misaligned gas disc through accretion, as can be seen in the third panel. This gas disc is seen to rotate in the fourth panel, due to continued misaligned gas accretion. At late times it exhibits a warp, clearly visible in the fifth panel, because the gas disc torques with the stars and the stellar torques are stronger in the centre, which therefore realigns first. The last panel shows that the gas disc is almost completely aligned with the stars. The gas disc remains misaligned for much longer than a few times $t_\mathrm{dyn}$.}
\end{figure*}

In the present work, we study the evolution of the gas disc embedded in the centre of a large early-type galaxy. Specifically, the galaxy experiences a gas-rich merger at $t_\mathrm{lookback}=3.4$~Gyr ($z=0.3$), which pushes out most of its gas disc. The original disc material reaccretes quickly, but is soon thereafter permanently removed from the system in a violent outflow. When the galaxy subsequently regrows a gas disc, this disc is initially misaligned with the stellar rotation, but eventually relaxes into the plane of the stars. 

To illustrate this evolution, we show the average line-of-sight velocity of all the gas in a (20~kpc)$^3$ box around the galaxy centre in Figure~\ref{fig:img} at several lookback times as indicated in each panel. The colour scale runs from $-300$~km~s$^{-1}$ (blue) to 300~km~s$^{-1}$ (red) and is saturated at the extremes. The image is rotated such that the stellar `disc' is edge on and horizontal, i.e. its angular momentum vector points in the $y$-direction. The shape of the stellar component is shown as white contours which connect areas of equal mass surface density, with logarithmically spaced contour levels. This also shows that the galaxy is an ETG during the entire evolution we consider here. At $t_\mathrm{lookback}\gtrsim4$~Gyr, the stellar axis ratio is reasonably small, resembling an S0. At later times, after the merger, the axis ratio is larger and the galaxy more resembles an elliptical.

The first panel shows the gas disc 0.7~Gyr before the merger, the second panel right after the merger when most of the gas disc has been removed. The galaxy reforms a gas disc through accretion, but it is misaligned with the stars, as can be seen in the third panel. The fourth panel shows that the disc tilts and precesses with respect to the stars, which is due to continued gas accretion as explained below. The central kpc precesses faster than the outskirts, starting a warp. At late times, it develops a warp in the outer parts as well, as shown in the fifth panel. Here, the evolution is driven by stellar torques, which are stronger in the central part than in the outer parts, resulting in the centre aligning first. By the end of the simulation, shown in the last panel, the gas disc is almost completely aligned with the stars.

\subsection{Evolution of galaxy properties} \label{sec:prop}

At $t_\mathrm{lookback}=5$~Gyr, the galaxy's stellar component shows a small disc structure in its kinematics, although it is clearly bulge-dominated. The stellar disc component disappears after the last major merger, but there is still a preferred rotation direction. The specific angular momentum decreases after the merger. We measure the mass-weighted 2D edge-on axis ratio, $b/a$, and $\lambda_R$, an observable proxy of the specific angular momentum \citep{Emsellem2007,Emsellem2011},
\begin{equation}
\lambda_R=\dfrac{R_\mathrm{2D}|v_\mathrm{LOS}|}{R\sqrt{v_\mathrm{LOS}^2+\sigma^2}},
\end{equation}
where $R_\mathrm{2D}$ is the half-mass radius measured in projection, $v_\mathrm{LOS}$ is the line-of-sight velocity, and $\sigma$ is the velocity dispersion in each (0.1~kpc)$^2$ pixel. We find that this galaxy is a fast rotator both before and after the merger, according to the classification of \citet{Emsellem2011}, $\lambda_R>0.31\sqrt{1-b/a}$. Before the merger, $\lambda_R=0.4-0.5$. Afterwards, $\lambda_R$ decreases, mostly due to the reaccretion of the stars removed in the merger, until it levels off at about 0.25 at $t_\mathrm{lookback}<2.6$~Gyr.
It is possible that the luminosity-weighted $\lambda_R$ would be even higher, since young stars have higher velocities, so the classification of this galaxy as a fast rotator is robust. The properties of the gas disc match observations of local ETGs well (see the discussion in Section~\ref{sec:concl}).

\begin{figure}
\center
\includegraphics[scale=.52]{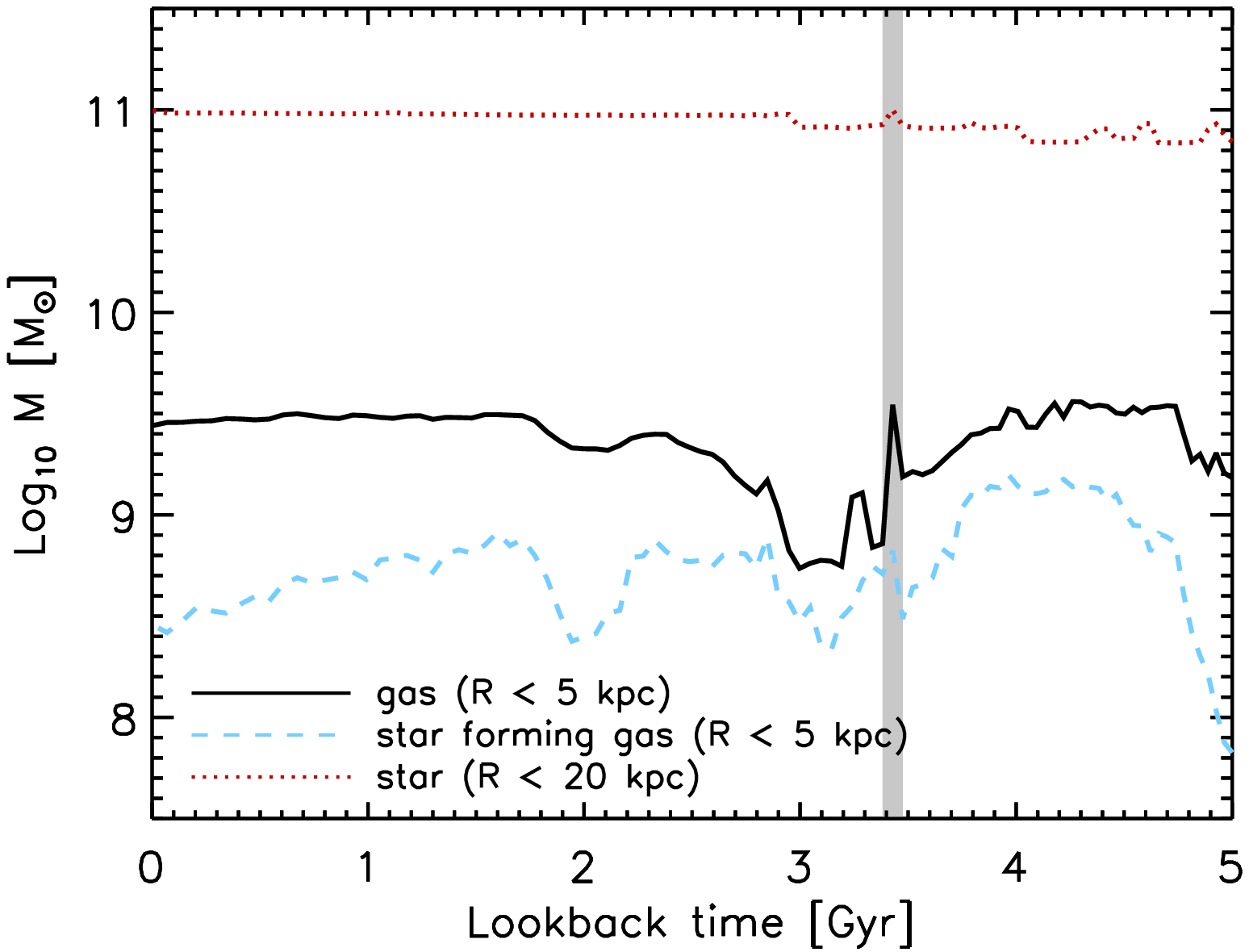}
\caption {\label{fig:mass} The total (solid, black curve) and star-forming (dashed, blue curve) gas mass within 5 kpc and stellar mass (dotted, red curve) within 20 kpc as a function of lookback time. The stellar mass is close to $10^{11}$~M$_\odot$ and only grows by a factor of 1.4 over the past 5~Gyr. The total gas mass is a few times $10^9$~M$_\odot$ for most of its recent history, but drops to $5\times10^8$~M$_\odot$ shortly after the merger (indicated by the grey band). The star-forming (i.e.\ dense) gas mass is highest before the merger (above $10^9$~M$_\odot$) and decreases to $2\times10^8$~M$_\odot$ after the merger. At late times, the total gas reservoir remains constant, but the star-forming gas reservoir decreases, due to star formation.}
\end{figure}

Figure~\ref{fig:mass} shows the total gas mass (solid, black curve) and the star-forming gas mass (dashed, blue curve), i.e. the gas at $n_\mathrm{H}>10$~cm$^{-3}$, in the central 5~kpc and the stellar mass (dotted, red curve) in the central 20~kpc. The stellar mass inside 5~kpc is only 0.2~dex lower than the total stellar mass shown. The mass of this galaxy is dominated by its stars within 25~kpc, outside of which dark matter dominates. At $z=0$, the stellar mass inside 20 and 5~kpc is $10^{11}$ and $6\times10^{10}$~M$_\odot$, respectively, and has only grown by a factor of 1.7 and 1.3 in the past 5~Gyr. In the central 5~kpc, about half the stellar mass increase is due to the accretion of stars originally belonging to the merging galaxy and the other half is due to star formation. The stellar half-mass radius at late time is 3.4~kpc for stars within 20~kpc.

The galaxy that merges and destroys most of the gas disc does so on its second passage. Just before the merger, the satellite's gas mass inside 5 kpc and stellar mass inside 20 kpc are $2.2\times10^9$ and $2.2\times10^{10}$~M$_\odot$, respectively. This is a factor of 1.4 larger than the central galaxy's gas mass and a factor of 3.8 smaller than the central galaxy's stellar mass at the same time. With a stellar mass ratio of 1:4 this merger is on the border of being classified as a major merger, whereas in gas mass the merger ratio is close to 1:1. 

The gas mass decreases by a factor of a few after the merger to $5\times10^8$~M$_\odot$. Most of the gas in the outer parts of the disc is removed, but the central kpc remains relatively intact. The half-mass radius of gas within 5~kpc is only 0.4~kpc right after the merger, but grows steeply after $t_\mathrm{lookback}=3$~Gyr, staying between 2.5 and 3.2~kpc after $t_\mathrm{lookback}=2.1$~Gyr. The total gas mass continues to grow down to $t_\mathrm{lookback}=1.7$~Gyr to a total of $3\times10^9$~M$_\odot$ and stays stable after that. The star-forming gas mass also decreases after the merger, but does not build back up as much as the total gas mass. Additionally, it decreases steadily after $t_\mathrm{lookback}=1.7$~Gyr, due to star formation and relatively low accretion rates. The star-forming gas mass is an order of magnitude lower than the total gas mass at the end of the simulation, whereas at e.g.\ $t_\mathrm{lookback}=4$~Gyr the difference was less than a factor of three. The half-mass radius of the star-forming gas varies between 0.2 and 0.9~kpc over the past 5~Gyr. These gas masses and radii are consistent with the range observed in ETGs \citep[e.g.][]{Young2011, Davis2011, Davis2013}.

\begin{figure}
\center
\includegraphics[scale=.52]{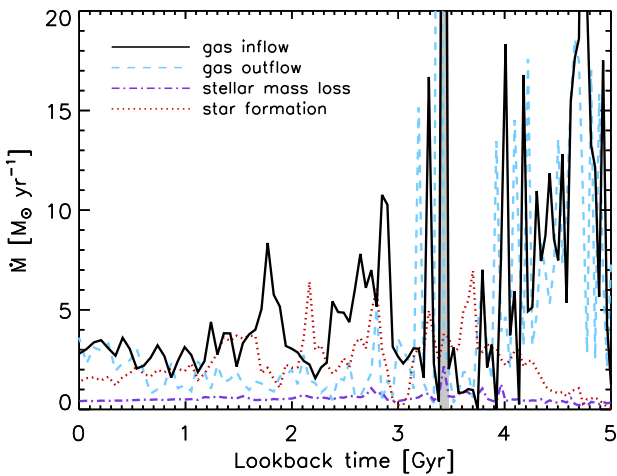}
\caption {\label{fig:rate} The gas inflow rate (solid, black), gas outflow rate (cyan, dashed), star formation rate (dotted, red), and stellar mass-loss rate (purple, dot-dashed) within 5~kpc from the centre as a function of lookback time. On average, the inflow rate dominates over the outflow rate and stellar mass-loss rate. After $t_\mathrm{lookback}=3.1$~Gyr, there are two significant periods of gas accretion, but there are no strong outflows. The gas accretion rate remains constant (and relatively low) at around 3~M$_\odot$~yr$^{-1}$ at $t_\mathrm{lookback}<1.5$~Gyr.}
\end{figure}

Inflow and outflow rates in the past 5~Gyr are shown in Figure~\ref{fig:rate} as the solid, black curve and dashed, light-blue curve, respectively. Additionally, the stellar mass-loss rate and star formation rate (SFR) are shown by the dot-dashed, purple curve and the dotted, red curve, respectively. All rates are measured within 5~kpc of the centre and averaged over the time between snapshots (which varies from 36 to 69 Myr). We checked that there is no significant difference when using a SFR averaged over 100 Myr instead. The stellar mass-loss rate shown here is defined as the increase in gas mass in the centre not due to particles leaving or entering the inner 5~kpc. It is much lower than the gas accretion rate and thus subdominant at reforming the gas disc and fueling star formation after the merger.   

The galaxy experiences strong inflow and outflow between $t_\mathrm{lookback}=4$ and 5~Gyr, which is due to close flybys by satellites and to a minor merger. The rates are much lower afterwards, but peak at the time of the main merger at $t_\mathrm{lookback}=3.4$~Gyr (at 48 and 59~M$_\odot$~yr$^{-1}$ for inflow and outflow, respectively) and shortly thereafter. After this time there are no major outflows. The galaxy experiences two episodes of relatively strong gas inflow (without associated strong outflow), peaking at $t_\mathrm{lookback}=2.9$~and 1.8~Gyr. The first peak is due to the accretion of the majority of the gas content of the galaxy that merges at $t_\mathrm{lookback}=3.4$~Gyr. The gas accretion rate remains low in the last 1.5~Gyr, consistent with the lack of late-time mass growth as seen in Figure~\ref{fig:mass}.

The SFR of this galaxy at $z=0$ is 1.4~M$_\odot$~yr$^{-1}$, which means its specific SFR is about $10^{-11}$~yr$^{-1}$, consistent with observed SFRs in gas-rich ETGs \citep[e.g.][]{Davis2014}. This galaxy lies below the blue cloud of star-forming galaxies \citep[e.g.][]{Brinchmann2004}, but always has a non-negligible gas reservoir and residual star formation (see Figures~\ref{fig:mass} and~\ref{fig:rate}). We also find a kinematically decoupled core in young stars, which are formed in the misaligned gas disc, as observed \citep{McDermid2006}. The SFR fluctuates between 0.1 and 7.0~M$_\odot$~yr$^{-1}$ over the last 5 Gyr. The average inflow and outflow rates over the past 3~Gyr, i.e. after the merger, are 3.8 and 1.7~M$_\odot$~yr$^{-1}$, respectively, and the average SFR is 2.4~M$_\odot$~yr$^{-1}$. The average stellar mass-loss is only 0.6~M$_\odot$~yr$^{-1}$ over the past 3~Gyr. 

Most of the gas that accretes onto the central 5~kpc in the past 5~Gyr does so in the `cold mode', i.e.\ it never reached temperatures close to the virial temperature of the halo ($T_\mathrm{vir}\approx1.4\times10^6$~K) in its history. In fact, only approximately 30 per cent of the gas that accretes after the merger ever had a temperature above $10^{5.5}$~K, a temperature at which the cooling time is relatively short \citep[see e.g.][]{Keres2005, Keres2009, Voort2011a, Faucher2011, VoortSchaye2012}. The majority of the gas that reforms the disc therefore did not cool from a diffuse, hot, hydrostatic halo, as was assumed in the semi-analytic model of \citet{Lagos2015}. Instead, this gas was brought in with high density by the merging galaxy and via satellites that were stripped in the halo, i.e.\ the gas accretion is not smooth, but clumpy. 

A substantial fraction of the gas originally in the disc is removed during the merger, but reaccretes after only about 0.1~Gyr. After another 0.1~Gyr a strong supernova-driven outflow occurs, reducing the gas mass once again, but the gas ejected in this event has not reaccreted by $z=0$. Only 40 per cent of the gas in the disc prior to the merger remains in the centre, forming stars. On the other hand, 2/3 of the gas brought in by the merging galaxy is accreted onto the central disc. Most of this accretion happens at $t_\mathrm{lookback}=2.9$~Gyr, substantially increasing the total gas mass in the disc.

\begin{figure}
\center
\includegraphics[scale=.52]{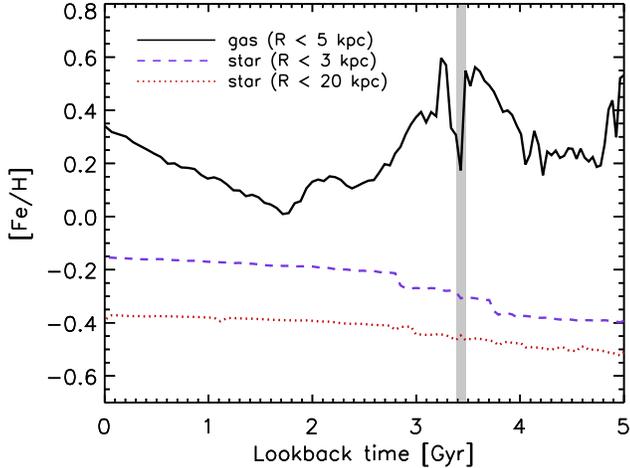}
\caption {\label{fig:FeH} Mean gas metallicity within 5~kpc (solid, black) and stellar metallicity within 20~kpc (dotted, red) and within 3~kpc (dashed, purple) of the galaxy centre. The decrease in gas metallicity after $t_\mathrm{lookback}=3$~and 2~Gyr corresponds to periods of high gas inflow rates, as seen in Figure~\ref{fig:rate}. This also shows that the galaxy is not reaccreting its pre-merger gas between $t_\mathrm{lookback}=3.2$ and 1.7~Gyr, but instead is accreting new material with lower metallicity. At later times, the gas metallicity increases again, because the accretion rate is low enough that the enrichment by supernovae dominates.}
\end{figure}

The importance of gas accretion is also obvious when we look at the evolution of the average metallicity, [Fe/H], shown in Figure~\ref{fig:FeH} for the gas within 5~kpc as a solid, black curve. [Fe/H] is the abundance ratio of iron to hydrogen compared to that of the Sun and is defined as $\mathrm{log}_{10}(N_\mathrm{Fe}/N_\mathrm{H})-\mathrm{log}_{10}(N_\mathrm{Fe}/N_\mathrm{H})_\odot$ where $N_\mathrm{Fe}$ and $N_\mathrm{H}$ are the number densities of iron and hydrogen. The stellar metallicity within 20~kpc and 3~kpc of the galaxy centre is shown by the dotted, red and dashed, purple curves, respectively. 3~kpc is close to the stellar half-mass radius. 

The stellar metallicity within 20~kpc (3~kpc) rises slowly from [Fe/H]$=-0.5$ ($-0.4$) to $-0.4$ ($-0.2$) by $z=0$. The slightly subsolar stellar metallicity and the fact that it decreases with increasing radius are both consistent with the majority of observed ETGs \citep[e.g.][]{Kuntschner2010}. The gas metallicity decreases after the merger from [Fe/H]=0.6 to 0 by $t_\mathrm{lookback}=1.7$~Gyr. This indicates that the accretion of fresh, metal-poor material dominates over the metals created by massive stars and expelled by stellar winds and is consistent with our finding that the gas mass increases substantially and gas accretion rates are high down to $t_\mathrm{lookback}\approx1.7$~Gyr (see Figures~\ref{fig:mass} and~\ref{fig:rate}). Afterwards, the metallicity increases again to [Fe/H]=0.3 at $z=0$, showing that at late times the metal dilution by accretion is less important than the metal enrichment by star formation.

\subsection{Misalignment and relaxation} \label{sec:align}

\begin{figure}
\center
\includegraphics[scale=.52]{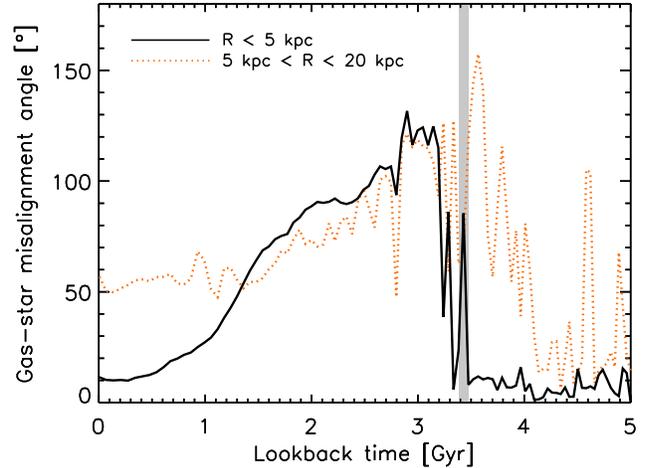}
\caption {\label{fig:angle} The misalignment angle between the angular momentum vector of the gas and stars as a function of lookback time is shown by the solid, black curve. A substantially misaligned disc is present for about 2~Gyr, much longer than a dynamical time. The misalignment angle for gas with radii between 5~and 20~kpc, i.e. gas which will likely accrete onto the gas disc, is shown by the dotted, orange curve. The inner and outer misalignment angles evolve together after the merger, until they diverge at $t_\mathrm{lookback}=1.3$~Gyr, after gas accretion slows down, enabling the stellar torques to realign the inner gas disc.}
\end{figure}

Figure~\ref{fig:angle} shows the misalignment angle between the angular momentum vector of the gas within 5~kpc and stars within 10~kpc as a function of lookback time as a solid, black curve\footnote{We use 10~kpc for stars instead of 20~kpc in order to be less affected by close satellite flybys, but the results are insensitive to this choice.}. The misalignment angle for gas with $R=5-20$~kpc, i.e. gas which is likely to accrete onto the gas disc in the near future, is shown by the dotted, orange curve. The central disc shows no misalignment before the merger. The gas at $R=5-20$~kpc starts to misalign at $t_\mathrm{lookback}=4$~Gyr, likely a result from gas stripping from close satellite flybys, but it does not affect the fairly massive central disc significantly. At the time of the merger, the disc loses most of its mass and subsequently accretes gas with misaligned angular momentum from the region around it. 
 
The gas at $R<5$~kpc and $R=5-20$~kpc show similar gas-star misalignment angles with the stars down to $t_\mathrm{lookback}=1.3$~Gyr. These angles decrease at the same rate and since the misalignment angle of the gas outside the disc is slightly lower than that of the disc, this also indicates that the continued accretion of gas is driving the change in the disc's angular momentum direction. At $t_\mathrm{lookback}<1.3$~Gyr, the central disc decouples and relaxes into the plane of the stellar `disc', but the gas at larger radii remains misaligned at a constant angle of about 55\degree. The gas at large radii cannot be driving the late time relaxation of the inner disc when the angle is less than 50\degree. It is possible that accreting gas, which has a larger misalignment angle, slows down the relaxation, but it will not be able to speed it up. 

The decoupling of the gas disc and the gas halo happens after the gas mass has stopped growing, the gas accretion rate has decreased, and the gas metallicity has started to increase again, as described in Section~\ref{sec:prop}. This is consistent with the picture that in the beginning of the evolution of the misaligned gas disc, the angular momentum brought in through gas accretion is high compared to the angular momentum change induced by stellar torques. In the later stages of the evolution, the gas accretion rate is lower and the mass of the disc is higher, resulting in gravitational torques dominating the change of the disc's angular momentum vector. We checked that the total angular momentum brought in by gas accretion is sufficient to explain the angular momentum change of the gas disc in the first 1.5~Gyr after the merger, but that it is not afterwards.

In the absence of gas accretion, this gas disc would relax to a misalignment angle of 180\degree, since it starts out at an angle larger than 90\degree. The fact that this does not happen again shows the important role of continued accretion from the gas around the central disc, which drives the misalignment angle to less than 90\degree. 

As explained in the introduction (Section~\ref{sec:intro}), one would naively expect relaxation into the plane of the stellar `disc' to take only a few dynamical times. However, we find that substantially misaligned gas disc is present for about 2 Gyr, which is much longer than a dynamical time, even at a radius of 5~kpc (see Figure~\ref{fig:relax} below). The persistence of the misalignment is mostly driven by continued misaligned accretion, at least down to $t_\mathrm{lookback}\approx1.3$~Gyr. 

The dark matter angular momentum in the central 20~kpc is misaligned with that of the stars by about 25\degree\ during most of the past 5~Gyr. The misalignment with the gaseous angular momentum vector is generally larger. There is no smooth trend as seen in Figure~\ref{fig:angle} and we emphasize that at these radii the stellar component dominates the potential and thus the torques on the gas. The dark matter at larger radii also exhibits large misalignment angles with both the gaseous and stellar angular momentum in the galaxy centre. We are therefore confident that the dark matter is not the main driver of the gas-star realignment.

\begin{figure}
\center
\includegraphics[scale=.52]{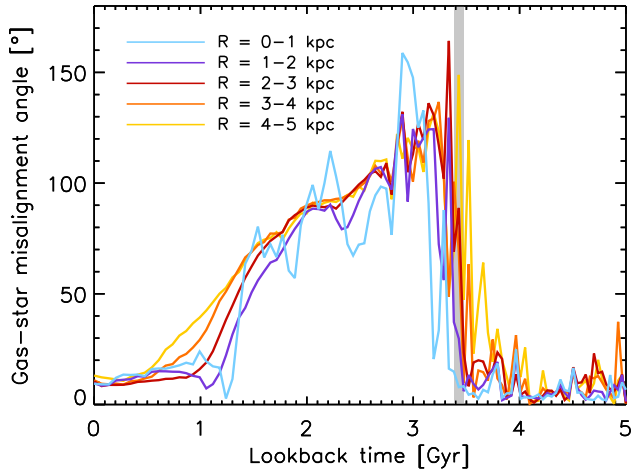}
\caption {\label{fig:shell} The misalignment angle between the gas and stellar angular momentum vectors as a function of lookback time as in Figure~\ref{fig:angle}, but dividing the gas disc into 5 shells of 1~kpc thickness. The shells at larger radii, $R$, become misaligned first around the time of the merger, while the inner kpc misaligns last. For the first part of the evolution ($t_\mathrm{lookback}\approx3-2$~Gyr), the shells have similar misalignment angles (and the gas disc is not warped). At $t_\mathrm{lookback}\lesssim1.9$~Gyr, the smaller the radius of the shell, the faster it relaxes into the plane of the stellar `disc', resulting in a warp.}
\end{figure}

Figure~\ref{fig:shell} shows the same misalignment angle as Figure~\ref{fig:angle}, with the difference that we subdivide the gas disc into spherical shells of 1~kpc thickness around the galaxy centre. All gas shells become misaligned with the stars around the time of the merger, but the shells at larger radii misalign first. At $t_\mathrm{lookback}\lesssim1.9$~Gyr, the smaller the radius of the shell, the faster it aligns its angular momentum vector with that of the stars. For example, the inner shell (light-blue curve) drops below 20\degree\ misalignment at $t_\mathrm{lookback}=1.3$~Gyr, while the outer shell (yellow curve) does so only at $t_\mathrm{lookback}=0.6$~Gyr. This causes the gas disc to develop a warp, as seen in the fourth and fifth panel of Figure~\ref{fig:img}. The warp is initially most pronounced in the centre and at late times in the outskirts. 

Especially the inner shell shows very non-monotonic behaviour, with the misalignment angle decreasing and increasing sharply. This indicates a strong role for gas accretion and feedback-driven outflows, since torquing with the stellar component will only decrease the misalignment angle. Another possibility is that the disc torques with satellites that pass at a short distance, but this would disrupt the outer disc more strongly than the inner disc. 

\begin{figure}
\center
\includegraphics[scale=.52]{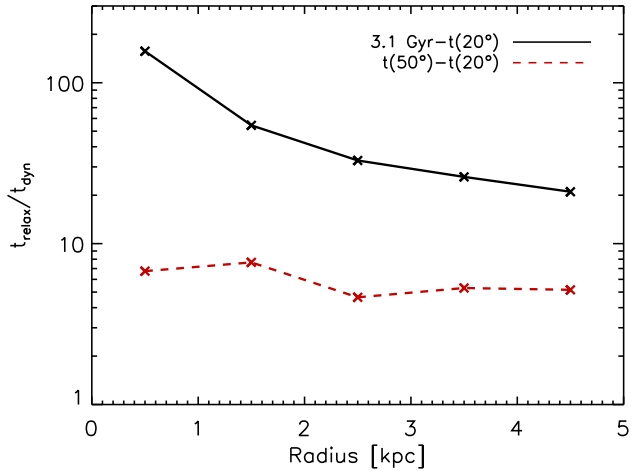}
\caption {\label{fig:relax} The relaxation time, determined in two different ways, of gas shells divided by the dynamical time at the middle of the shell as a function of the radius of the shell. The solid, black curve shows the total time the individual shells are misaligned compared to their dynamical time, i.e.\ starting at $t_\mathrm{lookback}=3.1$~Gyr (when all gas shells are strongly misaligned with the stars) and ending when the misalignment of an individual shell drops below 20\degree. The relaxation time defined in this way does not scale linearly with $t_\mathrm{dyn}$. The dashed, red curve shows the time it takes each shell to relax from 50\degree\ to 20\degree. 50\degree\ is the misalignment angle of the gas outside the disc (see Figure~\ref{fig:angle}), so below this value the relaxation cannot be driven by gas accretion. When defined in this way, $t_\mathrm{relax}$ is about 6~$t_\mathrm{dyn}$, as expected. This shows that initially the gas disc's evolution is driven by external accretion, whereas at late times gravitational torques dominate.} 
\end{figure}

We refer to the time it takes the gas to realign its angular momentum direction with that of the stars as the relaxation time, $t_\mathrm{relax}$. We investigate the relaxation of five 1~kpc thick gas shells around the galaxy centre. From Figure~\ref{fig:shell} it is clear that at $t_\mathrm{lookback}=3.1$~Gyr all gas shells have similarly large misalignment angles, of about 120\degree. We calculate $t_\mathrm{relax}$ from this time until the misalignment angle becomes less than 20\degree, when we consider the disc to be sufficiently relaxed. The solid, black curve  in Figure~\ref{fig:relax} shows this relaxation time (which varies from 1.8 to 2.5~Gyr) divided by the time-averaged dynamical time at half the shell radius, $t_\mathrm{dyn}=2\pi R_{1/2}/v_\mathrm{circ}$ (which varies from 0.01 to 0.12~Gyr), where $R_{1/2}$ is the radius at the middle of the shell and $v_\mathrm{circ}$ is the circular velocity at $R_{1/2}$. In the centre the misalignment is visible for over a hundred dynamical times and for several tens of dynamical times at larger radii. The relaxation time defined in this way does not scale linearly with $t_\mathrm{dyn}$.

If only gravitational torques were important for the relaxation of the gas disc, one would expect a constant scaling with the dynamical time, which is clearly not what we find in Figure~\ref{fig:shell}. However, as we showed above, in this cosmological simulation, it is not just the torques with the stars that are driving the evolution of the gas disc, but also the continued gas accretion from its surroundings. In Figure~\ref{fig:angle} we saw that the evolution of the misalignment angle of the gas outside the central disc decouples from that of the inner disc when it reaches about 50\degree. It is likely that afterwards the evolution of the misalignment is caused mostly by stellar torques (although external accretion could continue to slow down the relaxation). We therefore define $t_\mathrm{relax}$ of the individual gas shells in a second way by calculating the time it takes the misalignment angle to decrease from 50\degree\ to 20\degree. Note that this is a decrease in misalignment angle of only 30\degree\ and the `full' relaxation time would be at least a factor of 2 larger. This second relaxation time (which varies from 0.07 to 0.58~Gyr) divided by the dynamical time is shown as the dashed, red curve in Figure~\ref{fig:relax}. We find that $t_\mathrm{relax}/t_\mathrm{dyn}\approx5-8$ and there is no trend with radius, indicating that, in this case, $t_\mathrm{relax}$ scales linearly with the dynamical time, as expected \citep{Tohline1982, Lake1983}.

\section{Discussion and conclusions} \label{sec:concl}

We have shown results from a cosmological zoom-in simulation of an ETG, which experiences a 1:4 (1:1) merger in stellar mass (gas mass) at $t_\mathrm{lookback}=3.4$~Gyr ($z=0.3$). This merger destroys most of the central gas disc. We have studied the subsequent formation of a new gas disc, whose rotation direction is misaligned with the stellar rotation. We have quantified its evolution and, in particular, its eventual realignment with the galaxy's stellar angular momentum. 

The misaligned disc in our simulation matches the properties of observed misaligned gas discs. The star-forming gas mass is approximately 3$\times$10$^8$~M$_\odot$ at $z=0$. This gas is concentrated in the inner kpc and extends to about 1/3 of the half-mass radius of the galaxy. These properties are consistent with those observed for such ETGs \citep{Young2011, Davis2011, Davis2013}. The merger also produces a kinematically decoupled core of young stars in the galaxy centre, as is observed in some of these systems \citep[e.g.][]{McDermid2006}. 

Our results show that a misaligned gas disc can persist for several Gyr, much longer than its dynamical time (about 150~$t_\mathrm{dyn}$ in the central kpc). This effect is driven by the continued accretion of material from minor satellites, which dominates the angular momentum evolution of this system for the first $\gtrsim1.5$~Gyr after the initial (major) merger. Once the accretion rate has dropped sufficiently, gravitational relaxation of the gas disc does occur, over a timescale of about 6~dynamical times.

Polar disc galaxies, where an outer disc of gas and young stars is misaligned by about 90\degree\ from a smaller disc of old stars, and warps in outer gas discs have been studied in previous simulations \citep[e.g.][]{Brook2008, Roskar2010, Snaith2012}. These misaligned gas structures are found to form and persist for many Gyr through the continued accretion of cold gas with misaligned angular momentum, similar to the simulated ETG we discussed here.

\citet{Lagos2015} assume in their semi-analytic model that when the angular momentum of a dark matter halo is misaligned from the stellar angular momentum, a misaligned gas disc is present. Our results show that indeed gas (and dark matter) in the halo can be misaligned from the stars. However, the dark matter in our simulation is not aligned with the halo gas either. Furthermore, we showed that it is not just the gas-star misalignment angle of the accreting gas that matters, but also the rate at which mass (and thus angular momentum) is accreted to counteract stellar torques. Eventually, after the accretion rate decreases sufficiently, the gas disc relaxes into the stellar plane even though the halo gas is still misaligned. 

Davis \& Bureau (in preparation) use the observed misalignments of gas discs in ETGs to show that a long relaxation time is likely required to explain their misalignment distributions. In this work they postulate that a continued accretion of cold gas could slow relaxation sufficiently. Our results are consistent with this scenario and the relaxation timescale we find matches theirs closely. 

\citet{Serra2014} compare their H\,\textsc{i} measurements of local ETGs (on scales of tens of kpc) to cosmological simulations and find qualitative agreement in the cold gas kinematics, but too few of the simulated ETGs host misaligned gas discs. They speculate that a more sophisticated treatment of gas physics and feedback is needed in the simulations. It is possible that our simulation matches the observed sample more closely, since from Figure~\ref{fig:angle} we can see that the gas is misaligned out to at least 20~kpc for at least 4~Gyr, but a larger sample of simulated ETGs is necessary to be sure. 

We caution that from a single example we cannot draw strong conclusions about the evolution of misaligned gas discs in the general population of ETGs. The simulated galaxy we have studied in detail may not be representative for the behaviour of all ETGs. It sits in a dense group-like environment and is therefore not evolving in isolation. It has several satellites and it is 3.4~Mpc away from the centre of another $2\times10^{13}$~M$_\odot$ halo, also containing a group of galaxies. Galaxies in the field could have a very different accretion history. Studying high-resolution cosmological simulations of a large sample of ETGs in a variety of environments will be necessary to quantify how rare or common the evolution described in this paper is.

Low stellar surface brightness features of the galaxy merger, such as shells and tidal tails, are visible after the merger, but disappear before the gas disc aligns with the stellar component, consistent with merger visibility estimates \citep[e.g.][]{Bell2006}. Since the gas-star misalignment persists for so long, misaligned gas discs are not necessarily associated with observable merger signatures in the stellar photometry, even when the misalignment was originally caused by a merger. 

In our simulation, the misaligned gas disc has a single kinematic position angle for the first part of its life, i.e.\ the gas shells in Figure~\ref{fig:shell} all show similar misalignments for $t_\mathrm{lookback}\approx3.1-1.9$~Gyr. At later times, gravitational relaxation, which acts on a constant multiple of $t_\mathrm{dyn}$, causes the gas disc to warp once accretion has slowed sufficiently. \citet{Davis2011} did not find any evidence for warps in their observational sample of such discs, but with the low spatial resolution of their data it is not clear how strong a statement can be made. \citet{Serra2014} found warps in 35 per cent of ETGs, but on larger scales than studied here. Future searches for warps in a statistical sample of galaxies with high resolution data can help constrain the balance between gas accretion-driven misalignment (for which no warps are expected) and stellar torque-driven relaxation (for which we predict warped gas discs).

In order to complete such an analysis one requires 2D kinematic maps of gaseous and stellar tracers, such as those provided by integral field unit (IFU) spectrographs. Existing IFU surveys of galaxies are limited in size, but ongoing and future surveys, such as the Calar Alto Legacy Integral Field spectroscopy Area survey (CALIFA; \citealt{Garcia2014}), the Sydney-AAO Multi-object Integral field spectrograph galaxy survey (SAMI; \citealt{Bryant2015}), and the Mapping Nearby Galaxies at APO survey (MaNGA; \citealt{Bundy2015}), will contain hundreds to thousands of objects. These surveys will be able to make exquisite measurements of the kinematic misalignment distribution, and how it changes in different objects (e.g.\ with galaxy mass, environment etc.). Under the assumption that (major) mergers are responsible for creating misaligned gas discs and that we know how long the misalignments persist, these observations will become a powerful way to constrain the (major) merger rate, and its variation, in the nearby universe. If indeed continued gas accreting significantly affects the timescales over which misaligned discs remain observable, these surveys may teach us about cosmological gas accretion onto the ISM as well. Theoretical studies of the factors that affect the misalignment and realignment of gas discs in galaxies, using a large statistical sample of simulated ETGs, will be vital to understand the creation and persistence of misaligned gas discs and enable correct interpretation of these observations.

\section*{Acknowledgements}

We would like to thank the referee for valuable comments. FvdV would also like to thank Ann-Marie Madigan, Mariska Kriek, Chung-Pei Ma, and Sanch Borthakur for useful discussions. 
TAD acknowledges support from a Science and Technology Facilities Council Ernest Rutherford Fellowship.
DK was supported in part by NSF grant AST-1412153 and funds from the University of California San Diego.
EQ was supported in part by NASA ATP grant 12-APT12- 0183, a Simons Investigator award from the Simons Foundation, and the David and Lucile Packard Foundation.
CAFG was supported by NSF through grant AST-1412836 and by Northwestern University funds.
The simulation presented here used computational resources granted by the Extreme Science and Engineering Discovery Environment (XSEDE), which is supported by National Science Foundation grant number OCI-1053575, specifically allocations TG-AST120025 (PI Kere\v{s}), TG-AST130039 (PI Hopkins), TG-AST1140023 (PI Faucher-Gigu\`ere).

\bibliographystyle{mn2e}
\bibliography{misalign}

\bsp

\label{lastpage}

\end{document}